\documentclass[%
 reprint,
 amsmath,amssymb,
 aps,
prd,
]{revtex4-2}

\usepackage{graphicx}
\usepackage{dcolumn}
\usepackage{bm}

\begin{document}

\preprint{APS/123-QED}

\title{Pauli blocking effects on pair creation in strong electric field}
\author{Mikalai Prakapenia$^{1,2}$}
\author{Gregory Vereshchagin$^{2,3,4,5}$}
\affiliation{$^{1}$Department of Theoretical Physics and Astrophysics, Belarusian State
University, Nezalezhnasci Av. 4, 220030 Minsk, Belarus}
\affiliation{$^{2}$ICRANet-Minsk, Institute of Physics, National Academy of Sciences of
Belarus\\
220072 Nezalezhnasci Av. 68-2, Minsk, Belarus}
\affiliation{$^{3}$ICRANet, 65122 Piazza della Repubblica, 10, Pescara, Italy}
\affiliation{$^{4}$ ICRA, Dipartimento di Fisica, Sapienza Universit\`a di Roma, Piazzale Aldo Moro 5, I-00185 Rome, Italy}
\affiliation{$^{5}$INAF -- Istituto di Astrofisica e Planetologia Spaziali, 00133 Via del Fosso del Cavaliere, 100, Rome, Italy}
\date{\today}

\begin{abstract}
The process of electron-positron pair creation and oscillation in uniform electric field is studied, taking into account Pauli exclusion principle. Generally, we find that pair creation is suppressed, hence coherent oscillations occur on longer time scales. Considering pair creation in already existing electron-positron plasma we find that the dynamics depends on pair distribution function. We considered Fermi-Dirac distribution of pairs and found that for small temperatures pair creation is suppressed, while for small chemical potentials it increases: heating leads to enhancement of pair creation.
\end{abstract}

\maketitle



\section{Introduction}

Quantum electrodynamics predicts the creation of electron-positron pairs in
strong electric field out of vacuum as a nonperturbative process with the
field strength exceeding the critical value \cite{1951PhRv...82..664S} $%
E_{c}=m^{2}c^{3}/e\hbar\sim10^{18}$ V/m,\ where $m$ and $e$ are electron
mass and charge, respectively, $c$\ is the speed of light and $\hbar$ is
reduced Planck's constant. It was predicted more than ninety years ago \cite%
{1931ZPhy...69..742S}, right after the invention of positron by Paul Dirac 
\cite{1928RSPSA.117..610D,1931RSPSA.133...60D}. So far this process is not
observed in the laboratory, despite strong efforts in increasing electric
field strength, in particular by focusing ultraintense optical laser beams,
see e.g. \cite{2012RvMP...84.1177D,2022arXiv220300019F}. Due to copious
amount of pairs created in such electric field it is believed that the
process of pair creation cannot be considered on a fixed background,
so that accounting for back reaction of newly created particles on the
external field is mandatory in this problem.

The study of the process of electron-positron pair creation and oscillations
induced by back reaction in the homogeneous time dependent electric field
has a long history, for reviews see \cite{Ruffini2009,2023PhR..1010....1F}. Comparison of
solutions of quantum Vlasov equations with classical kinetic
Vlasov-Boltzmann equations performed in \cite%
{1991PhRvL..67.2427K,1992PhRvD..45.4659K,1998PhRvD..58l5015K} showed that
classical description is in surprisingly good agreement with the full
quantum treatment even for the field strengths $E>E_c$. It was also shown
that for largely overcritical fields quantum treatment leads to
non-Markovian kinetic equations \cite%
{1998IJMPE...7..709S,1998PhRvD..58l5015K}. Numerical solutions of these equations were obtained in \cite{1999PhRvD..59i4005S}, showing that for overcritical field memory effects become important. Finally, effects of quantum statistics were analyzed in \cite{2021PhRvD.104c1902N} and shown to be important when quantum interference occurs \cite{2022Symm...14.2444A}. A hydrodynamic approach developed
in \cite{2007PhLA..371..399R,2011PhLB..698...75B} following \cite%
{1987PhRvD..36..114G} allowed to establish that plasma oscillates with a
frequency comparable to the plasma frequency. The study of this problem with
Boltzmann-Vlasov equations describing, in addition to pair creation, also
interaction with photons \cite{2013PhLA..377..206B} showed that plasma
thermalization occurs on much longer timescales than oscillations do, see also \cite{1999PhRvD..60k6011B,2007PhRvL..99l5003A}.%

Pair production in strong electric field has been also discussed in the
context of early Universe, in particular during inflation \cite%
{2019PhRvD.100l3502G}. The role of the gravitational field in the process of
pair creation is considered in \cite{2018PhRvD..98d5015F}. Of course, one of
the key directions of these efforts is its verification in laboratory
experiments \cite{2022PhRvL.129x1801K}.

Very recently \cite{2021PhRvD.104c6007F} quantum Vlasov
equations were derived from the nonequilibrium quantum field theory. Backreaction of electron-positron pairs onto rapidly
oscillating electric field was studied in \cite{2023arXiv230104026J} using
quantum Vlasov equations \cite{2014PhRvD..90l5033H} confirming that only for
undercritical field the backreaction can be safely neglected. The validity
of locally constant field approximation is discussed in \cite%
{2021PhRvD.104g6014S}. 

Recently Pauli blocking effects in thermalization of relativistic plasma
were studied in \cite{2019PhLA..383..306P,2020PhLA..38426679P,2022Univ....8..473V}. So far no systematic analysis of the influence of quantum degeneracy on pair creation and plasma oscillations was carried out. Our previous works did not include the effects of Pauli blocking. In this work we close this gap.

The paper is organized as follows. In Section 2 the framework is presented:
Boltzmann-Vlasov equations for pairs together with Maxwell equation for electric field. In Section 3 main results for vacuum initial state are reported. In Section 4 non-vacuum initial state is considered and the role of inverse Schwinger process is emphasized. Conclusions follow in the last section.

\section{Framework}

There are two main assumptions in this work. Firstly, following the results in \cite{1991PhRvL..67.2427K,1992PhRvD..45.4659K,1998PhRvD..58l5015K} we assume that classical kinetic equations provide good approximation to quantum dynamics of pairs created in overcritical electric field. Secondly, we assume that pair creation rate computed for vacuum state does not change when electron-positron pairs are present.

In this section we present a kinetic description based on the relativistic Boltzmann-Maxwell equation with the source term accounting for pair creation in strong electric field \cite{1987PhRvD..36..114G}, modified to include the Pauli blocking effect. In what follows the system of units $\hbar = c = 1$ is adopted, then $e=\sqrt{\alpha}$, where $\alpha$ is the fine structure constant.

As uniform electric field $E(t)$ is considered, the problem has axial symmetry.
We introduce cylindrical coordinates in momentum space $\mathbf{p}%
=\{p_{\perp },\phi ,p_{||}\}$ with the $p_{||}$-axis parallel to electric
field $E$. Particle energy is then $p^{0}=[p_{\perp
}^{2}+p_{||}^{2}+m^{2}]^{1/2}$.

Particle evolution is described by one-particle electron/positron
distribution function $f(t,p_{\perp },p_{||})$, which is normalized on
particle density $n=\int \frac{d^{3}p}{(2\pi )^{3}}f$. Energy density (of electrons and positrons) and energy per particle are defined as follows:\ $\rho \ =2\int d^{3}p(2\pi )^{-3}p^{0}f$ and $\epsilon =n^{-1}\int \frac{d^{3}p}{(2\pi )^{3}}p^{0}f$. Since for this work there is no difference between electrons and positrons (apart from the opposite direction of acceleration by the electric field) in the following for definiteness we denote $f$ the positron distribution function. Particle collisions are neglected, as they occur on much larger time scales than what is considered in this work, leading eventually to plasma thermalization \cite{2013PhLA..377..206B}.

Collisionless Boltzmann equation governing evolution of $f(t,p_\perp,p_{||})$
is, see e.g. \cite{1991PhRvL..67.2427K,1992PhRvD..45.4659K}: 
\begin{equation}
\frac{\partial f}{\partial t} + e E \frac{\partial f}{\partial p_{||}} =
S(E,p_\perp,p_{||}),  \label{eq1}
\end{equation}
where $S$ is the source term for Schwinger process: 
\begin{gather}
S=-(1-2f) e|E| \nonumber \\ \times\text{ln}\left[1-\exp\left(-\frac{\pi(p_\perp^2 + m^2)}{e|E|}%
\right) \right] \delta(p_{||}),  \label{eq2}
\end{gather}
with the factor $(1-2f)$ being the Pauli blocking accounting for both electrons and positrons.

Time evolution of electric field is defined from the Maxwell equation $\frac{%
dE}{dt}=-j_{\text{cond}}-j_{\text{pol}}$ containing conductive current $j_{%
\text{cond}}$ generated by the motion of pairs and polarization current $j_{%
\text{pol}}$ generated by the pair creation process. These currents are
defined as follows: 
\begin{gather}
j_{\text{cond}}\ =2e\int \frac{d^{3}p}{(2\pi )^{3}}\frac{p_{||}}{p^{0}}%
f, \\ j_{\text{pol}}\ =2e|E|E^{-1}\int \frac{d^{3}p}{(2\pi )^{3}}p^{0}S,
\label{eq3}
\end{gather}%
where factor 2 is included to account for both electrons and positrons. Then the
Maxwell equation becomes: 
\begin{gather}
\frac{dE}{dt}=-e\int \frac{d^{3}p}{(2\pi )^{3}}\frac{p_{||}}{p^{0}}f~+\frac{e|E|}{E}
\label{eq4} \\
\times\int \frac{d^{3}p}{(2\pi )^{3}}(1-2f)\text{ln}\left[ 1-\exp
\left( -\frac{\pi (p_{\perp }^{2}+m^{2})}{|qE|}\right) \right] p^{0}\delta
(p_{||}).  \notag
\end{gather}

It is useful to introduce dimensionless quantities $\tilde{t}=tm$, $\tilde{p}%
=p/m$, $\tilde{E}=Ee/m^{2}$ and rewrite eqs. (\ref{eq1}) and (\ref{eq4}%
) in dimensionless form: 
\begin{gather}
\frac{\partial f}{\partial \tilde{t}}+\tilde{E}\frac{\partial f}{\partial 
\tilde{p}_{||}}= ~ \label{eq5}  \\
-(1-2f)|\tilde{E}|~\text{ln}\left[ 1-\exp \left( -\frac{\pi (\tilde{p}%
_{\perp }^{2}+1)}{|\tilde{E}|}\right) \right] \delta (\tilde{p}_{||}), \notag
\end{gather}
\begin{gather}
\frac{\partial \tilde{E}}{\partial \tilde{t}}=-2e^{2}\int \frac{d^{3}\tilde{p%
}}{(2\pi )^{3}}\frac{\tilde{p}_{||}}{\tilde{p}^{0}}f+ \frac{2e^{2}|E|}{E}(1-2f)~ \label{eq6} \\
\times\int \frac{d^{3}\tilde{p}}{(2\pi )^{3}}\text{ln}%
\left[ 1-\exp \left( -\frac{\pi (\tilde{p}_{\perp }^{2}+1)}{|\tilde{E}|}%
\right) \right] \tilde{p}^{0}\delta (\tilde{p}_{||}), \notag
\end{gather}%
here $\tilde{p}^{0}=\sqrt{\tilde{p}_{\perp }^{2}+\tilde{p}_{||}^{2}+1}$ and $d^{3}\tilde{p}=2\pi \tilde{p}_{\perp }d\tilde{p}%
_{\perp }d\tilde{p}_{||}$ is the phase space element.

Equations (\ref{eq5}) and (\ref{eq6}) are solved numerically using the finite
difference scheme. For this goal we define a grid in the $\{\tilde
p_\perp,\tilde p_{||} \}$ space as follows. The $\tilde p_\perp$-grid is
logarithmic containing 10 nodes covering the interval $(0.001,50)$. The $%
\tilde p_{||}$-grid is uniform containing 500 nodes covering the interval $%
(-1000,1000)$. Replacing momentum derivatives by
finite differences in the Boltzmann equation (\ref{eq5}) we use upwinding
scheme for both positive and negative values of electric field $\tilde E$.
Then on a finite grid the Boltzmann equation (\ref{eq5}) transforms into the
system of ordinary differential equations for time variable $\tilde t$.
The integral in the r.h.s. of equation (\ref{eq6}) transforms to a finite sum. We use the implicit Gear's method to solve the system of ordinary differential
equations.

\section{Vacuum initial state}

We explore the process of pair creation in overcritical electric field with
different initial conditions for electric field and pairs. First we focus on
overcritical electric field with vacuum initial state. Then we turn to
initial state with electron-positron pairs distributed according to the
Fermi-Dirac statistics in an overcritical electric field.

\begin{figure}[tbp]
\includegraphics[width=\columnwidth]{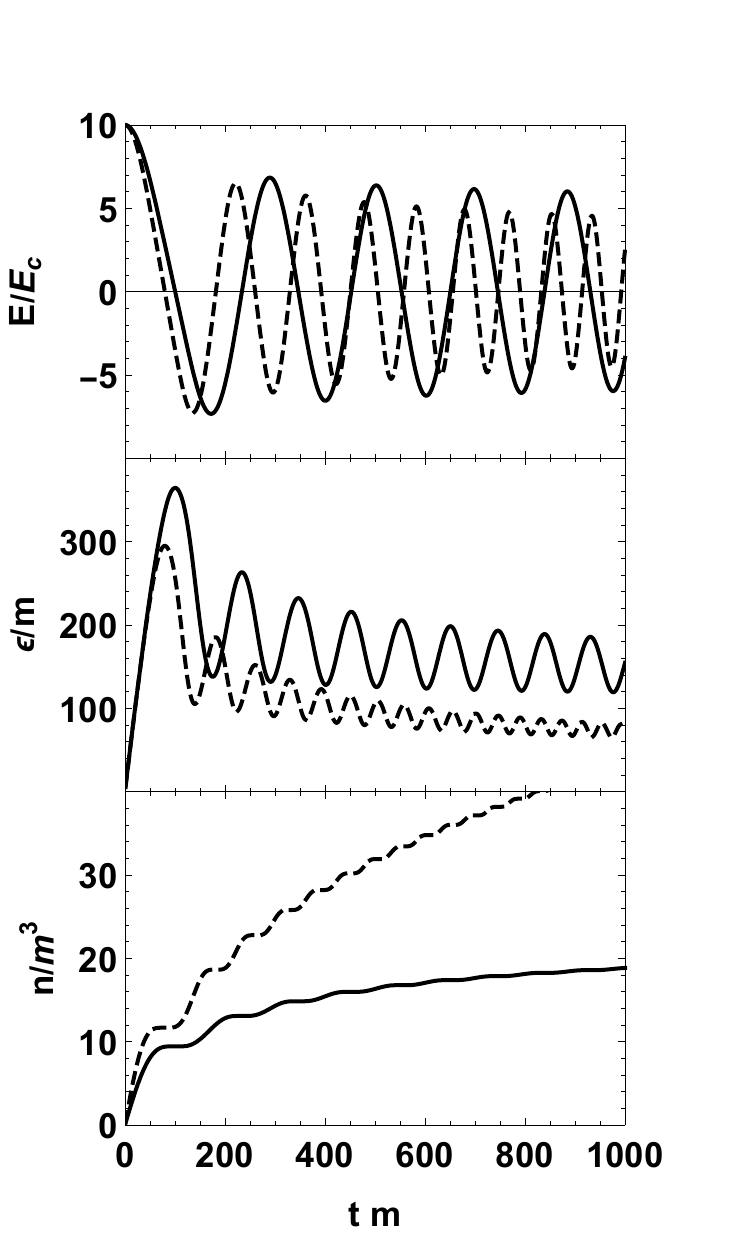}
\caption{Time evolution of electric field $E$, positron number density $n$
and average energy per particle $\protect\epsilon $ for $E_{in}=10E_{c}$.
Pauli blocking factor included on solid curve and excluded on dashed curve.}
\label{fig1}
\end{figure}

\begin{figure}[tbp]
\includegraphics[width=\columnwidth]{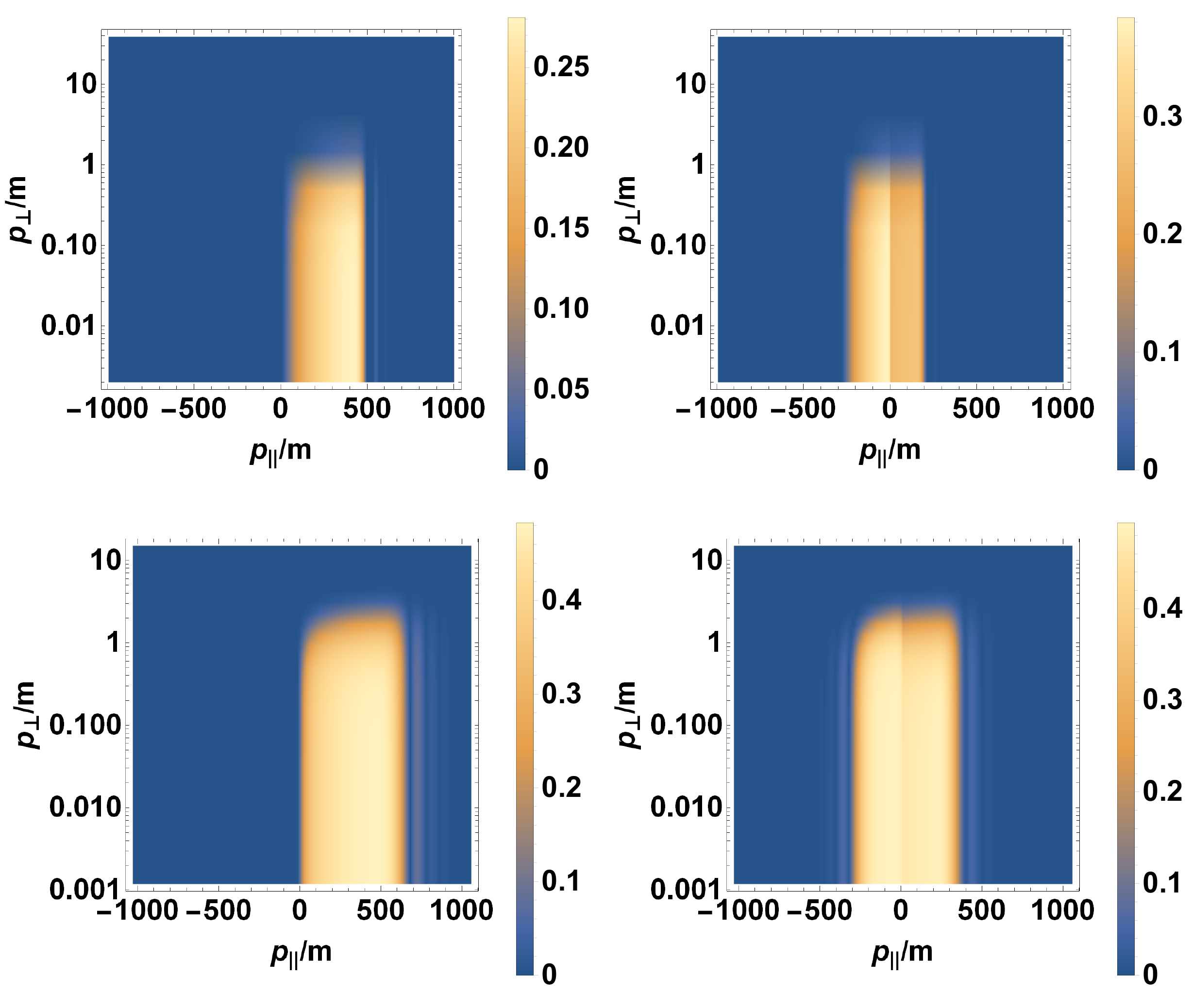}
\caption{Positron distribution function $f$ for the first oscillation at two
time moments: when electric field equals zero (left column) and when
electric field acquires local minimum (right column). Upper row correspond
to $E_\text{in}/E_c=3$. Lower row correspond to $E_\text{in}/E_c=11$.}
\label{fig2}
\end{figure}

In Fig. \ref{fig1} we demonstrate the effect of Pauli blocking on the pair
creation process by comparing time evolution of dimensionless quantities:\
electric field $\tilde{E}=E/E_{c}$, positron number density $\tilde{n}%
=n/m^{3}$ and average energy per particle $\tilde{\epsilon}=\epsilon /m$ in
two cases: when Pauli blocking is accounted for (solid curves), and when it
is neglected (dashed curves). In both cases plasma oscillations develop due to backreaction of the pairs: electric current is induced due to charged particle acceleration in the electric field; then particles overshoot thanks to their inertia and change the direction of the field. This process repeats as damped oscillations, due to creation of new pairs. It is clear that when Pauli blocking is taken into account the rate of pair creation is strongly suppressed and consequently oscillation frequency is smaller, and particle average energy is higher. It implies that oscillations are damped on a longer time scale. We also illustrate in Fig. \ref{fig2} the\ phase space evolution for selected initial electric field strengths. Comparing the upper and lower figures with different initial electric fields we find that pairs are produced with zero parallel momentum and with orthogonal momentum up to the value $p_{\perp }\sim \sqrt{E}$. Once pairs are created they are accelerated in the direction parallel to the electric field and hence are distributed over the parallel momentum $p_{||}$ rather uniformly. The degree of degeneracy is different:\ the larger electric field, the higher it is. Comparing the left and the right columns in Fig. \ref{fig2} we find that before electric field vanishes particles are gaining positive $p_{||}$, while after this moment particles are directed towards negative $p_{||}$. In the process of particle motion electric field continues to create new pairs:\ particle density is higher with negative $p_{||}$ than it is with positive $p_{||}$ on the right column.

Next, we study the pair creation process as function of initial electric
field $E_{in}$. Dimensionless energy density for electric field is $\tilde{%
\rho}_{E}=\tilde{E}^{2}/(2\alpha )$ and for pairs it is $\tilde{\rho}_{\text{pair}%
}\ =2\tilde{\rho}$. For vacuum initial state the energy conservation gives $\tilde{\rho}_{E}+\tilde{\rho}_{\text{pair}}=\tilde{\rho}_{E_{in}}$. The amount of energy transferred into pairs can be determined as $%
\tilde{\rho}_{\text{pair}}\ /\tilde{\rho}_{E_{in}}=1-\tilde{\rho}_{E}/\tilde{%
\rho}_{E_{in}}$. In Table \ref{tab1} we summarize some characteristic
quantities for pair creation process from vacuum depending on initial
electric field $E_{in}$: $\tilde{t}_{0}$ is time moment when electric field
vanishes for the first time; $\tilde{n}(\tilde{t}_{0})$ and $\tilde{\epsilon}%
(\tilde{t}_{0})$ are positron density and average energy in this moment; $%
\tilde{t}_{1/2}$\ is time moment when $f$\ increases to the value $1/2$; $\tilde{n}(\tilde{t}_{1/2})$ and $\tilde{\epsilon}(\tilde{t}%
_{1/2})$ are positron density and average energy in this moment; $\tilde{\rho%
}_{\text{pair}}\ (\tilde{t}_{1/2})/\tilde{\rho}_{E_{in}}$ is relative fraction
of the energy transferred into pairs at this moment.\ The main results are:\
with increasing initial electric field both the frequency of oscillations
and pair density monotonically increase. Average energy per particle first
decreases then, starting at about $6E_{c}$ it increases, see \cite%
{2011PhLB..698...75B,2013PhLA..377..206B}.\ For $E_{in}>10E_{c}$ the distribution function of pairs overcomes the value $1/2$ before the moment $\tilde{t}_{0}$ and we do not report the quantities at $\tilde{t}_{1/2}$. Pauli blocking
operates in such a way that $f=1/2$ is reached. Average energy per particle at the moment $\tilde{t}_{1/2}$\
monotonically increases as particles occupy more and more orthogonal
momentum space.

\begin{table}[tbp]
\centering%
\begin{tabular}[t]{|c|c|c|c|c|c|c|c|}
\hline
$\tilde{E}_{in}$ & $\tilde{t}_{0}$ & $\tilde{n}(\tilde{t}_{0})$ & $\tilde{%
\epsilon}(\tilde{t}_{0})$ & $\tilde{t}_{1/2}$ & $\frac{\tilde{\rho}_{\text{%
pair}}\ (\tilde{t}_{1/2})}{\tilde{\rho}_{E_{in}}}$ & $\tilde{n}(\tilde{t}%
_{1/2})$ & $\tilde{\epsilon}(\tilde{t}_{1/2})$  \\ \hline\hline
1 & 1431 & 0.057 & 600 & 88070 & 0.64 & 0.31 & 72   \\ \hline
3 & 264 & 0.99 & 311 & 1010 & 0.55 & 1.76 & 99   \\ \hline
6 & 136 & 4.04 & 306 & 204 & 0.6 & 4.35 & 171   \\ \hline
10 & 100 & 9.43 & 365 & 130 & 0.84 & 9.59 & 300   \\ \hline
10.1 & 99.8 & 9.58 & 366 & 1.53 &  &  &    \\ \hline
30 & 49 & 65.23 & 631 & 0.25 &  &  &    \\ \hline
60 & 33.7 & 198.15 & 839 & 0.11 &  &  &    \\ \hline
100 & 27.6 & 353.47 & 971 & 0.10 &  &  &    \\ \hline
1000 & 8.7 & 11302 & 3048 & 0.01 &  &  &    \\ \hline
\end{tabular}%
\caption{Results for pairless initial state.}
\label{tab1}
\end{table}

\section{Non-vacuum initial state}

Below we present the results of simulations with non vacuum initial state. The statistical factor $(1-2f)$ in equations (\ref{eq5}) and (\ref{eq6}) plays a crucial role in interaction between electric field and pairs. For initial conditions with $f<1/2$ pair creation from electric field leads to increase in particle number density and damping of oscillations. This effect is well described in the literature, see e.g. \cite{1991PhRvL..67.2427K,2011PhLB..698...75B,2013PhLA..377..206B}.

On the contrary, for $f>1/2$ the statistical factor becomes negative, which implies negative source term in equation (\ref{eq5}) and also opposite sign of the polarization current in equation (\ref{eq6}). Under these conditions the inverse Schwinger process, namely \emph{pair annihilation in external electric field} takes place. Quantum electrodynamics predicts that the rate of the inverse Schwinger process is equal to the rate of the direct one. This effect is clearly absent in vacuum and to our knowledge it is not discussed in the literature so far.

As we are interested in the effects of quantum degeneracy, we explore the influence of particle distribution in the phase space on dynamics of pairs and electric field. Electron-positron pairs in initial state are assumed to obey the Fermi-Dirac statistics
\begin{equation}
f=\left[ 1+e^{\left(\sqrt{\tilde{p}^{2}+1}-\tilde{\mu}\right)/\tilde{T}}\right] ^{-1},
\label{FDdistr}
\end{equation}%
where $\tilde{T}=T/m$ is dimensionless temperature and $\tilde{\mu}=\mu /m$
is dimensionless chemical potential. Note that equilibrium distribution with relativistic temperature and $\mu=0$ corresponds to $f<1/2$, while fully degenerate distribution with $T=0$ corresponds to $f>1/2$.



\subsection{Non-vacuum initial state with $f<1/2$}

\begin{figure}[tbp]
\includegraphics[width=\hsize,clip]{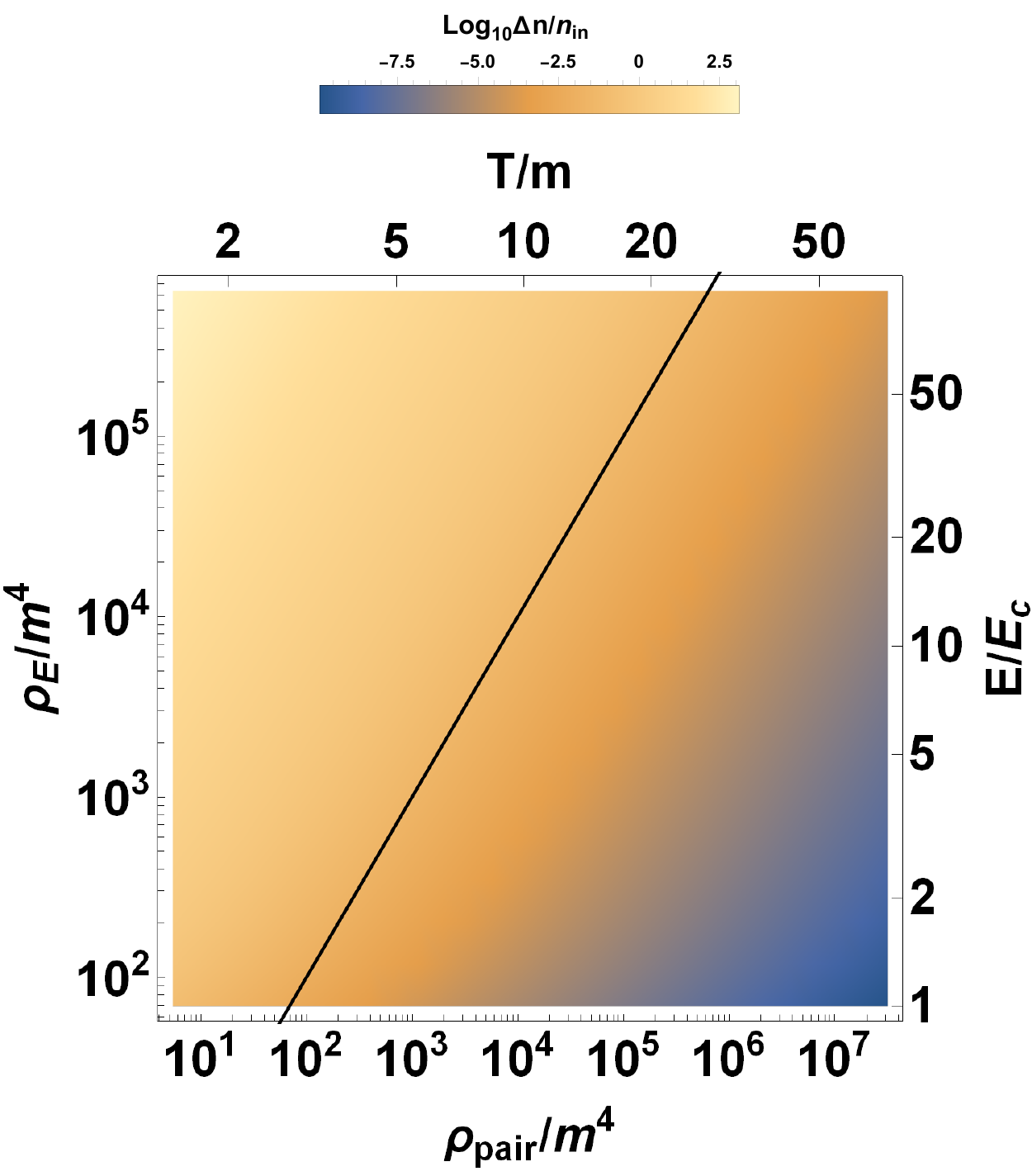} 
\caption{Relative change of pair number after 3 oscillation periods depending on initial electric field strength and plasma temperature. We also indicate energy density of electric field and pairs. Black line corresponds to equality $\rho_{pair}=\rho_E$. Chemical potential is zero.}
\label{fig3}
\end{figure}

In this section we consider initial distribution function of pairs with $f<1/2$. First we treat the case of distribution (\ref{FDdistr}) with $\mu=0$. In Fig. \ref{fig3} we show  the relative number of pairs produced after three oscillations depending on initial electric field and pair temperature. The dynamics in this case is qualitatively similar to the case with vacuum initial state discussed above. From Fig. \ref{fig3} it is clear that in electric energy domination region (above black line) pair production is efficient. Conversely, in the pair energy domination region (below black line) pair production is suppressed.
This is expected, because in the pair dominated region the Schwinger process is suppressed and plasma keeps oscillating with relativistic plasma frequency $\tilde{\omega}_p\ =\sqrt{\alpha \tilde{n}_{\text{pair}}/\tilde{\epsilon}}$. When initial conditions are in the electric field dominated region (black line in Fig. \ref{fig3}), the field accelerates particles much stronger. As electric field drags particles out from the region $p_{||}=0$ the phase space opens up and pair creation becomes possible.

In general, both temperature and chemical potential in (\ref{FDdistr}) can be nonzero. As we are interested in the influence of particle distribution function on the Schwinger process we consider two types of initial conditions: those with the same energy density and those with the same number density. The first choice corresponds to the same point on diagram in Fig. \ref{fig3} but with different parameters of the distribution function (\ref{FDdistr}). The second choice allows exploration of the role of heating. In fact, when number density is kept constant and the temperature increases, the average energy per particle increases. On diagram in Fig. \ref{fig3} it represents a shift to the right.

\begin{figure}[tbp]
\includegraphics[width=\hsize,clip]{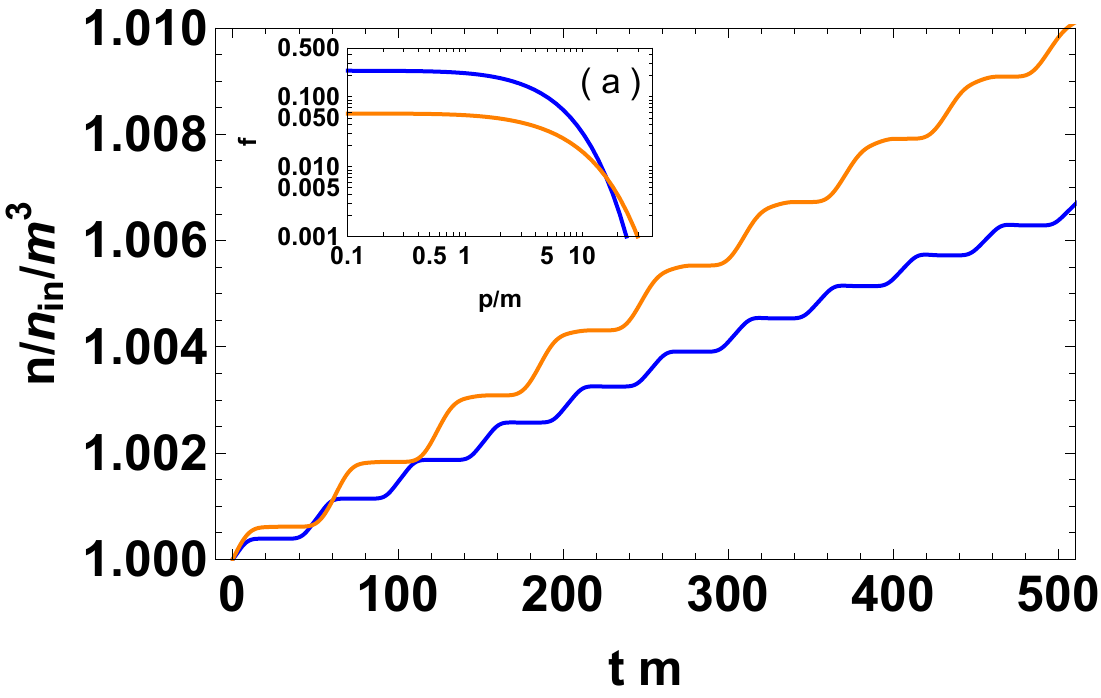} 
\caption{Time evolution of pair number density with initial electric field $E_{in}=E_c$ and two different initial pair states. Blue: $\tilde\mu=-3.71,\tilde T=4;$ orange: $\tilde\mu=-18.57,
\tilde T=7;$ Inset(a): the corresponding initial distribution functions.}
\label{fig4}
\end{figure}

In Fig. \ref{fig4} we show relative number of pairs with the same initial number density $n/m^3 = 9.66$, but different initial energy density $\rho/m^4 = 119.7$ (blue curve) and $\rho/m^4 = 204.4$ (orange curve), evolving in initial electric field $E_{in}=E_c$. Pair initial distributions are shown on the inset.
While pair creation is small in both cases, it clearly increases for initial conditions with higher energy density. This demonstrates the effect of heating of initially present plasma onto the pair creation process. The result is that despite the energy density of pairs increases due to heating, and hence initial conditions shift onto the pair dominated region in Fig. \ref{fig3} disfavoring pair creation, the effect of the opening up of the phase space due to the change of the distribution function prevails and pair creation becomes enhanced.

\subsection{Non-vacuum initial state with $f>1/2$}

\begin{figure}[tbp]
\includegraphics[width=\hsize,clip]{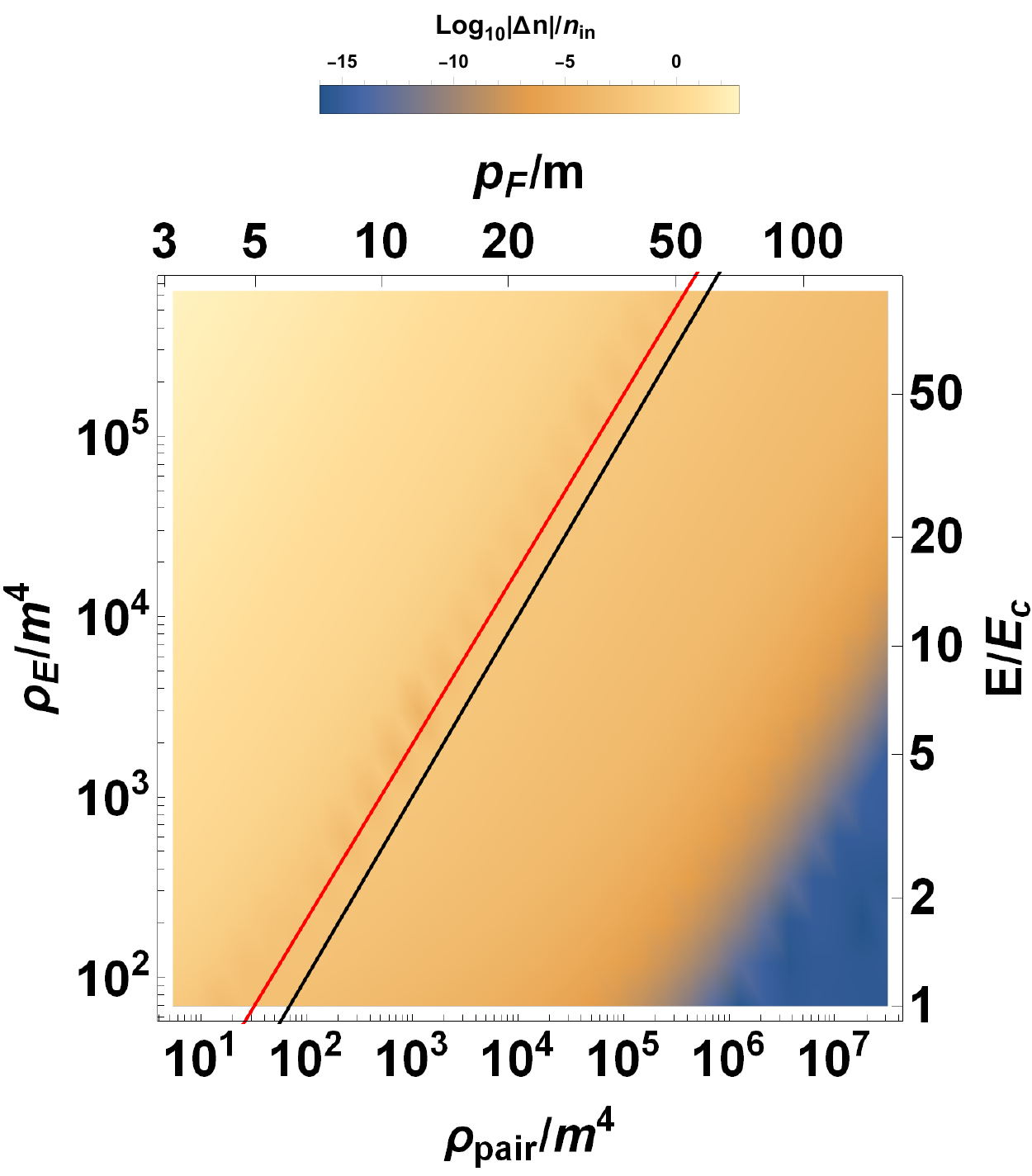}
\caption{The absolute value of the relative change of pair number after 3 oscillation periods depending on initial electric field strength and plasma chemical potential. We also indicate energy density of electric field and pairs. Black line corresponds to equality $\rho_{pair}=\rho_E$. Red line corresponds to transition from pair creation (above this line) to pair annihilation (below this line). For convenience we plot here Fermi momentum $p_F$ instead of the chemical potential. Temperature is zero.}
\label{fig5}
\end{figure}
In this section we consider initial distribution function of pairs (\ref{FDdistr}) with $T=0$. In Fig. \ref{fig5} we show the absolute value of the change in relative number of pairs after three oscillations depending on initial electric field and pair chemical potential or equivalently Fermi momentum $p_F$. In contrast with the previous case, here $f>1/2$ and statistical factor in equations (\ref{eq5}) and (\ref{eq6}) become negative, which implies particle annihilation in external electric field. During time evolution the number density of pairs increases above the red line and decrease below it.
Naively one could expect that pair annihilation would result in amplification of electric field as the process, opposite to pair creation and field depletion, shown in Fig. \ref{fig1}. However, there is no direct analogy in this process. Diminishing of the number of pairs (and hence their rest mass energy) leads not to increase of the energy density of electric field, but to increase of internal energy of pairs. This is because electric field accelerates particles redistributing them in momentum space. There is no possibility to use inverse Schwinger process to enhance electric field.
\begin{figure}[tbp]
\includegraphics[width=\hsize,clip]{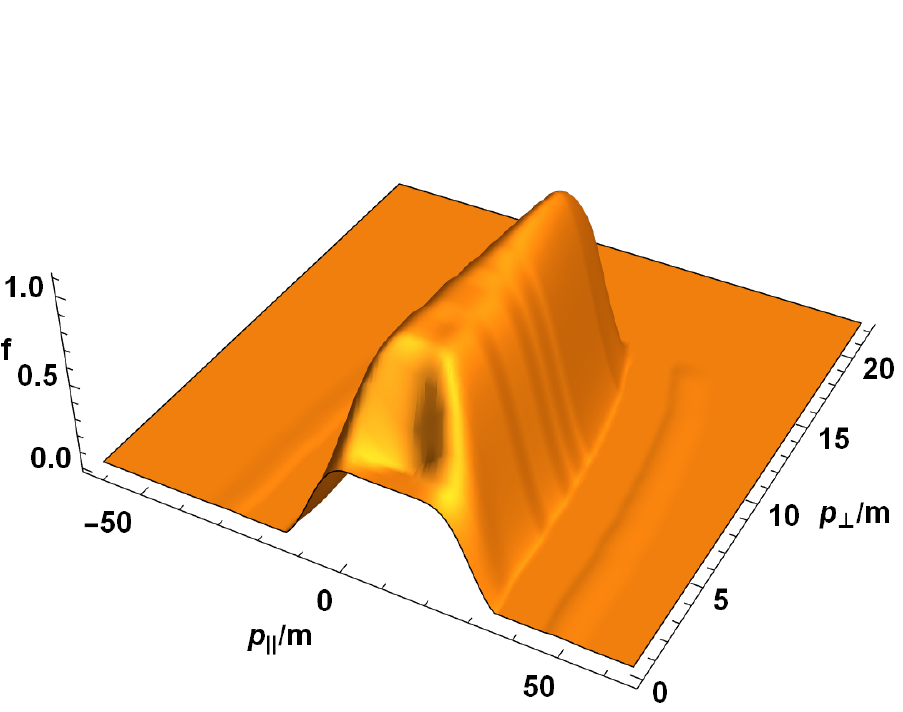} 
\caption{Distribution function of electron-positron pairs after 16 oscillations with $E=4E_c$ and initial distribution (\ref{FDdistr}) with $T=0$ and $\mu=16$.}
\label{fig8}
\end{figure}

In Fig. \ref{fig8} we present the distribution function $f(p_\perp,p_{||},t)$ after 16 oscillations. Note different scale in momentum axes. It is evident that pair annihilation leads to depletion of the distribution function only for small $p_\perp$, where the source term is significant. For larger $p_\perp$ the source term is negligible. As the source term is largest in absolute value just below the red line in Fig. \ref{fig5}, this imposes a limit on the effect of pair annihilation because the energy density of electric field cannot exceed much the energy density of pairs. In other words, the effect can be enhanced by increasing initial electric field, but this leads also to increase of the Fermi momentum in initial distribution function, thus reducing the part of the distribution affected by pair annihilation.

\begin{figure}[tbph!]
\includegraphics[width=\hsize,clip]{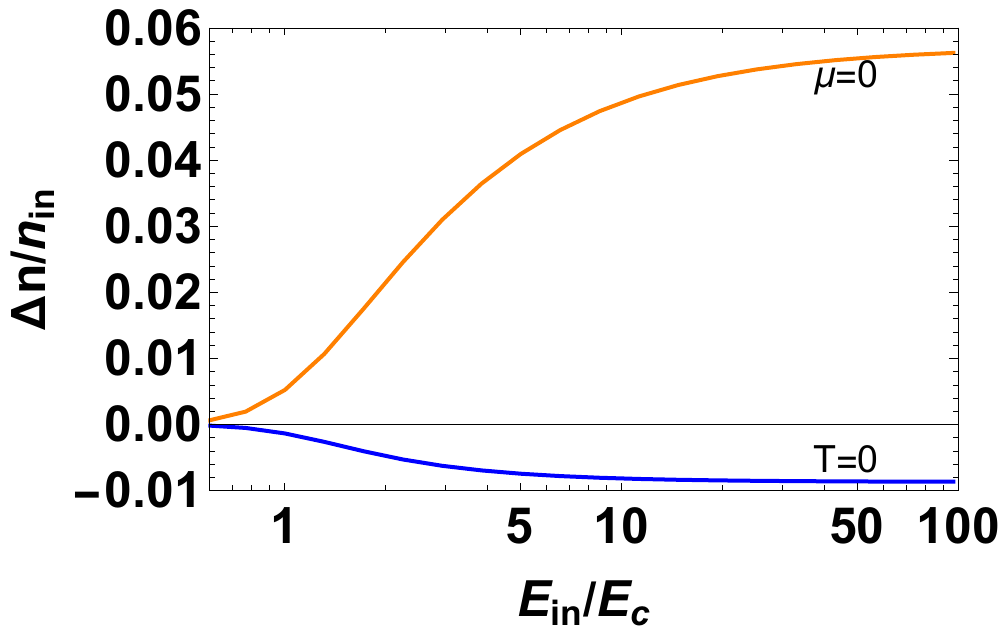}
\caption{Relative change of pair number after 3 oscillations
as function of electric field strength when initial electric field energy
equals initial pair energy. Blue curve correspond to fully degenerate pairs
with $T=0$ and orange curve correspond to $\mu=0$.}
\label{fig6}
\end{figure}
\begin{figure}[tbph!]
\includegraphics[width=\hsize,clip]{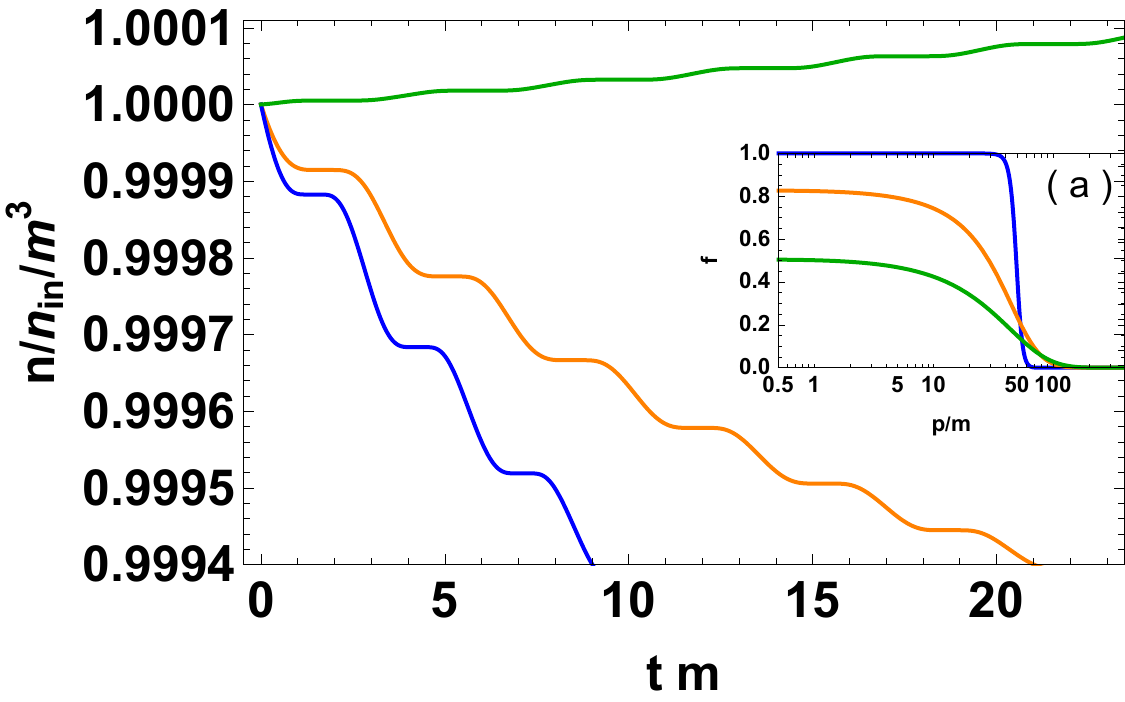} 
\caption{Time evolution of pair number density with initial electric field $%
E_{in}=10E_c$ and three different initial pair states. Blue: $\tilde\mu=49.41,\tilde T=3;$ orange: $\tilde\mu=29.21,
\tilde T=18;$ green: $\tilde\mu=1.58,\tilde T=28.$ Inset(a): the corresponding initial distribution functions}
\label{fig7}
\end{figure}
In Fig. \ref{fig6} we show the change in the relative number of pairs after three oscillations as function of initial electric field for two cases with the same initial energy density of pairs: pairs with zero chemical potential and relativistic temperature (orange curve) and fully degenerate pairs with zero temperature (blue curve). Clearly the first case corresponds to $f<1/2$ and leads to pair creation, while the second case represents initial conditions with $f>1/2$ and leads to pair annihilation. Both cases show saturation at large electric fields. All initial conditions represented in Fig. \ref{fig5} are located on the black line in Fig. \ref{fig3}.



In Fig. \ref{fig7} we present time evolution of pair number density with $E_{in}=10E_c$ and with initial pairs number density $\tilde n_{in}=4.2\times 10^3$ for three different initial pair states: $\tilde\mu=49.41,\tilde T=3$ ($\tilde\rho=1.6\times 10^5$); $\tilde\mu=29.21, \tilde T=18$ ($\tilde\rho=2.6\times 10^5$) and $\tilde\mu=1.58,\tilde T=28$ ($\tilde\rho=3.7\times 10^5$). The inset in Fig. \ref{fig6} illustrates the distribution function for these initial conditions. As can be seen, only the case with smallest chemical potential corresponds to $f<1/2$ and pairs are created; in other two cases pairs annihilate with time.

\section{Conclusions}

Our main result in this work is the demonstration how quantum exclusion principle suppresses pair creation in overcritical uniform electric field, which in turn modifies the back reaction dynamics. We studied electron-positron pair creation and oscillations with initial vacuum state as well as with electron-positron plasma initially present. Two cases can be distinguished. 1) When the energy in electric field dominates that in pairs oscillations are induced, which leads to opening up of the phase space and consequent prolific pair creation. 2) In the opposite case, when pairs dominate energetically over electric field, plasma oscillations do occur with much higher frequency, since electric field is unable to displace them significantly in momentum space: as a consequence pair creation remains strongly suppressed.

We also considered the effect of the inverse Schwinger process, namely annihilation of pairs in external electric field when the statistical factor becomes negative. Despite naive expectation that pair annihilation could lead to amplification of electric field, we found that this is not the case, even for the limiting case of completely degenerate initial distribution function with $T=0$. Despite the number of pairs may significantly decrease, backreaction of pairs on electric field leads to transformation of their rest mass energy into their internal energy, and not the energy of electric field.

We found that plasma heating leads to enhancement of pair creation. This effect may be relevant for astrophysical models of quark stars or neutron stars with strong electric field on their surface \cite{1998PhRvL..80..230U,2012NuPhA.883....1B}.

After this paper has been submitted for publication we learned about very similar work \cite{2023PhRvE.107c5204B} just published. Using different formalism the authors of this publication reached similar conclusions and their results are consistent with ours. However, it appears they did not look into the evidence for the inverse Schwinger process.

{\bf Acknowledgements.} This work is supported within the joint BRFFR-ICRANet-2023 funding programme under the grant No. F23ICR-001. We are grateful to Alexander Fedotov for the discussions on the inverse Schwinger process. We also appreciate the comments of the anonymous referees, which allowed to improved the paper.

\bibliographystyle{unsrt}

\end{document}